\documentclass[aps,pre,preprint,superscriptaddress,showpacs,letterpaper]{revtex4-1} %2-column: twocolumn (alternativ: p5, preprint)

\usepackage[english]{babel}
\usepackage{graphicx}
\usepackage{geometry}
\usepackage{hyperref}
\usepackage{amsmath}
\usepackage{amssymb}
\usepackage{array}

\newcommand\eqnref[1]{(\ref{#1})}
\newcommand\figref[1]{Fig.~\ref{#1}}

\newcommand\sectref[1]{Section~\ref{#1}}

\newcommand{\bfH}   {\mathbf{H}}
\newcommand{\bfB}   {\mathbf{B}}
\newcommand{\bfE}   {\mathbf{E}}
\newcommand{\bfD}   {\mathbf{D}}
\newcommand{\bfb}   {\mathbf{b}}
\newcommand{\bfd}   {\mathbf{d}}
\newcommand{\bfe}   {\mathbf{e}}
\newcommand{\bfh}   {\mathbf{h}}

\newcommand{\bfr}   {\mathbf{r}}
\newcommand{\bfv}   {\mathbf{v}}

\newcommand{\bfw}   {\mathbf{w}}

\begin{document}

%\preprint{}

\title{Effective Constitutive Parameters of Plasmonic Metamaterials:\\
A Direct Approach}

\author{Anders Pors}
\email[Corresponding author: ]{pors@mci.sdu.dk}
%\thanks{}
\affiliation{Mads Clausen Institute (MCI), University of Southern Denmark, Alsion 2, DK-6400 S{\o}nderborg, Denmark}

\author{Igor Tsukerman}
\affiliation{Department of Electrical and Computer Engineering, The University of Akron, OH 44325-3904, USA}

\author{Sergey I. Bozhevolnyi}
%\email[]{seib@iti.sdu.dk}
\affiliation{Institute of Technology and Innovation (ITI), University of Southern Denmark, Niels Bohrs All\'{e} 1, DK-5230 Odense M, Denmark}

\date{\today}

%Abstract
\begin{abstract}
We introduce a general implementation of the recently proposed homogenization theory
[Tsukerman, J. Opt. Soc. Am. B 28, 577 (2011)] allowing one to retrieve 
all 36 linear constitutive parameters of any 3D metamaterial with parallelepipedal unit cells. 
The effective parameters are defined directly as linear relations between pairs of
coarse-grained fields, in contrast with methods where these parameters are obtained
from reflection/transmission data or other indirect considerations.
The method is applied to plasmonic metamaterials
with spherical gold particles and split-ring resonators (SRR), respectively.
In both cases, the expected physical behavior is reproduced almost perfectly, with no unphysical
artifacts. 
%With regard to the effective parameters of the SRR-metamaterial, comparison of the new
%approach with the existing ones reveals that effective parameters are, in general,
%weakly influenced by the choice of the method, due to the approximations involved.
\end{abstract}

%\pacs{78.20.-e} % Optical properties of bulk materials/thin films
 % Optical constants
%\pacs{78.20.Bh} % Methods, theory, numerical simulation in optics
%\pacs{81.05.Xj} %Metamaterials
%\pacs{73.20.Mf} %Plasmones on surfaces
%\pacs{41.20.Jb} % Electromagnetic wave propagation
%\pacs{42.25.-p} % Wave optics
%\pacs{75.30.Kz} %Magnetic phase boundaries (metamagnetism)
%\pacs{75.75.-c} %Magnetic properties of nanostructures
%\keywords{}

\maketitle

\section{Introduction}
In recent years, very extensive research effort has been devoted to investigating
and designing new artificial materials, called metamaterials, with staggering
new optical properties. Some spectacular phenomena realized with metamaterials
are artificial magnetism \cite{linden}, negative refraction \cite{shelby,shalaev},
super resolution \cite{fang}, electromagnetic cloaking \cite{schurig} and electromagnetically induced transparency \cite{zhang,papasimakis,liu,bozhevolnyi}.
Although these properties are not found in nature, metamaterials are in fact
made of conventional materials that are structured on a scale smaller
than the free-space wavelength $\lambda_0$ of light: the unit cell $a$ of metamaterials is typically
in the range $a \sim 0.1 - 0.4 \lambda_0$.
Consequently, one may look for effective material parameters describing
the optical properties of such nanostructured composite materials on the macroscale.
The definition and computation of such effective constitutive parameters is, however,
not a trivial task and in general involves numerical simulation,
as accurate analytical characterization of the electromagnetic fields
and parameters can only be obtained in special cases,
e.g. for a lattice of subwavelength spheroidal particles \cite{maxwell,doyle,lewin}
or simplified circuit-like models of split-ring resonators (SRR) \cite{marques}.

The literature on homogenization of metamaterials is now so vast that we shall
not attempt to review it here. Instead, the key distinguishing features
of our approach are highlighted below. Several monographs \cite{choy,sarychev,cai} 
and review papers \cite{simovski,tretyakov} are available;
see also \cite{fietz2} and \cite{tsukerman}.

In the following we would like to comment on conceptual differences between parameter retrieval methods and first-principles derivation.
The linear relation between the four coarse-grained electromagnetic fields $\bfE$, $\bfH$, $\bfD$, $\bfB$
-- defined by proper averaging \cite{tsukerman} of the ``microscopic'' fields $\bfe, \bfd, \bfb$ --
can, in general, be represented by a $6\times 6$ matrix that is usually (and especially for
plasmonic materials) frequency-dependent.
There are two general routes for finding the entries of this matrix.
In the first one the effective parameters of a material slab are inferred from the $S$-parameters,
i.e. from reflection and transmission of a plane wave through that slab
\cite{smith,chen,li,kwon,linden,yan,bozhevolnyi,chen2}.

The second route of analysis involves the fundamental definition of effective parameters
as relationships between the coarse-grained (``macroscopic'') fields \cite{smith2,fietz,tsukerman}.
To this end, the coarse-grained fields must be clearly and unambiguously defined
and linear relationships between them found.

It is important to appreciate that the first route -- parameter ``retrieval'' --
is, despite its great practical significance, fundamentally different from
the second one and should not be viewed from the same perspective.
A simple analogy may help to clarify the difference. Molecular properties
of a given chemical substance can be either inferred from spectroscopic measurements
or, alternatively, obtained from an \emph{ab initio} quantum-mechanical analysis.
Clearly, these two approaches differ substantially -- they can and should go hand-in-hand
but each of them is important in its own right.

The methodology proposed in \cite{tsukerman} and implemented in this paper
is akin to the first-principles quantum-mechanical analysis in the analogy above.
These first principles are Maxwell's equations as well as 
\emph{the very definition of effective parameters
as relations between (pairs of) coarse-grained fields} inside the material.
In the absence of this fundamental definition, it would be truly surprising 
that the effective parameters measured for slabs of different thickness
(let alone different shapes) would come out in agreement, even for natural materials
in classical electrodynamics. The very consistency of effective parameters derived
from different measurements and different geometries stems from the underlying
fundamental physical characteristic -- namely, again, effective parameters as relations
between coarse-grained fields.

Thus one distinguishing feature of our approach is that it is \textit{direct} --
the effective parameters are not inferred from some ancillary data but
are derived from the fundamental definition. This way, all 36 material parameters
are properly and uniquely defined and can be found.
In contrast, the ``retrieval'' procedures derive the effective parameters
indirectly and allow one to determine only a subset of these parameters.
Being in essence an inverse problem, parameter retrieval may be ill-posed
and may suffer from the multiplicity of solutions due to the branch
ambiguities that must be properly handled \cite{chen2}.

Another key distinguishing feature of our methodology has to do with the way
the coarse-grained fields are defined \cite{tsukerman}.
In a significant, but well justified, departure from the traditional ways 
of field averaging, we require that the coarse-grained fields be constructed 
to satisfy exactly Maxwell's boundary conditions at all interfaces --
namely, the tangential continuity of the $\bfE$ and $\bfH$ fields and the normal continuity
of $\bfD$ and $\bfB$ fields. Violation of these boundary conditions would be equivalent 
to spurious surface currents and/or charges, with the related nonphysical artifacts. 
For metamaterials, standard volume averaging (or, more generally, averaging by convolution 
with a smooth Gaussian-like mollifier) is not adequate in this regard. 
Indeed, if such averaging is applied within each of two abutting materials separately, 
then nonphysical field jumps will generally occur at the interface. 
If, alternatively, the moving average is applied across the interface, 
the fields will be smeared in a boundary layer, which is again nonphysical. 
Such spurious effects are negligible when the lattice cell size is vanishingly 
small relative to the wavelength (which is usually the case for natural materials), 
but for metamaterials nonphysical conclusions may result 
because the cell sizes are appreciable.

Thus the microscopic fields, in our approach, are subject to two different types of interpolation.
The $\bfE$ and $\bfH$ fields are obtained from $\bfe$ and $\bfb/\mu_0$
by an interpolation that preserves tangential continuity,
while $\bfD$ and $\bfB$ must be obtained from $\bfd$ and $\bfb$
by an interpolation that preserves normal continuity \cite{tsukerman}.
The fact that $\bfh$ (i.e., $\bfb/\mu_0$) and $\bfb$ must be subject to different types of interpolation
explains, among other things, why the macroscopic $\bfH$ and $\bfB$ fields may be different
even though their microscopic counterparts are the same (for intrinsically nonmagnetic media);
this paradox is at the heart of ``artificial magnetism'' \cite{pendry, shalaev2}.

Although the theory of \cite{tsukerman}, reviewed in the following section,
uniquely defines the full $6\times 6$ parameter matrix, it has so far been applied
to several test cases involving a subset of the 36 parameters \cite{tsukerman}.
This paper describes a general implementation of this method in 3D for a parallelepipedal unit cell,
for any type of inclusions and without any \textit{a priori}
assumptions about the effective parameters. The implementation is described in \sectref{sec:method}.
Sec.~\ref{sec:testcases} is devoted to instructive test cases of plasmonic metamaterials
with gold spherical particles and split-ring resonators (SRR). The metamaterial with a lattice of spheres
was chosen because Lewin's theory \cite{lewin} is available for comparison.
The SRR-metamaterial is interesting because of its ubiquity and the bianisotropic response \cite{chen,xu}.
In Sec. \ref{sec:compare}, the computed parameters of the SRR-metamaterial are compared
with the parameters obtained by the S-parameter methods. Conclusions are given in Sec. \ref{sec:conclusion}.

%%%%%%%%%%%%%%%%%%%%%%%%%%%%%%%%%%%%%%%%%%%%%%%%%%%%%%%%%%%%%
\section{The Method and Its Implementation}\label{sec:method}
%%%%%%%%%%%%%%%%%%%%%%%%%%%%%%%%%%%%%%%%%%%%%%%%%%%%%%%%%%%%%
%
\subsection{Vectorial Interpolation for the Coarse-Grained Fields}\label{sec:Vectorial-interpolation}
In this section, we implement the direct definition of parameters  outlined in the Introduction.
Namely, the coarse-grained fields are rigorously defined and a linear relationship between them established.
As already emphasized, the interpolation procedures for constructing the ``macroscopic''
fields must be chosen carefully, so that the normal or tangential continuity of the respective
fields across all interfaces is honored.

\begin{figure}
	\centering
	\includegraphics[width=8cm]{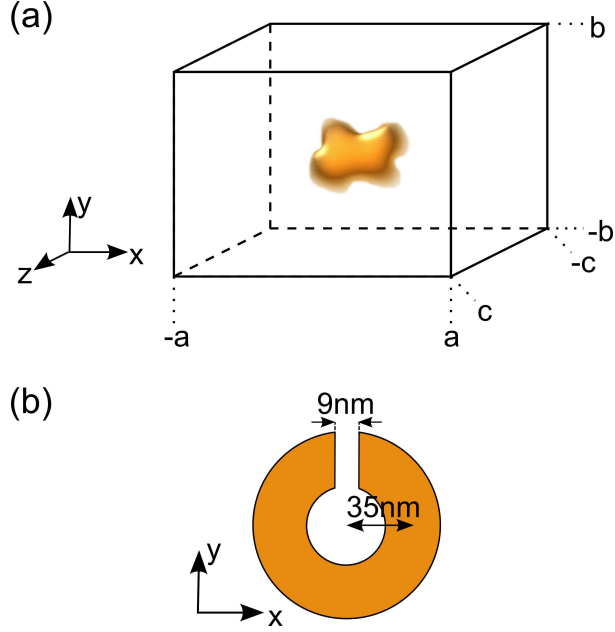}
	\caption{a) Sketch of the unit cell of size $2a\times 2b\times 2c$ containing
     an arbitrary metallic inclusion. b) Cross section of a split-ring resonator (SRR) lying in the $xy$-plane with a bending radius and gap of 35~nm and 9~nm, respectively. The width and thickness of the SRR are also 35~nm.}
	\label{fig:setup}
\end{figure}

Even though the relevant theory was already presented in \cite{tsukerman}, it is revisited here,
to make the paper more easily readable and self-contained. Some details and implementation
issues that have not been previously discussed are also presented here.
We assume that the unit cells are parallelepipeds containing arbitrary inclusions
[Fig.~\ref{fig:setup}(a)] with a linear complex dielectric
permittivity $\epsilon(\bfr, \omega)$. 
The size of the unit cell is $2a\times 2b\times 2c$
along the $x, y, z$ directions, respectively.

\begin{figure}
  \centering
  \includegraphics[width=8cm]{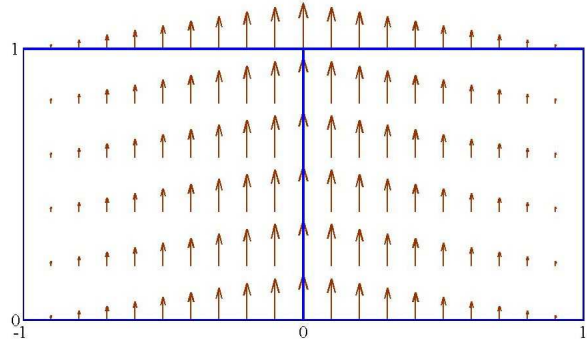}\\ %0.9\linewidth
  \caption{(From \cite{tsukerman}.) A 2D analog of the vectorial interpolation function $\bfw_\alpha$
  (in this case, associated with the central vertical edge shared by two adjacent cells).
  Tangential continuity of this function is evident from the arrow plot;
  its circulation is equal to one over the central edge and to zero
  over all other edges.}\label{fig:vector-edge-function}
\end{figure}

The interpolation procedure that preserves tangential continuity is effected by
vectorial interpolation functions like the one shown in \figref{fig:vector-edge-function},
in a 2D rendition for simplicity. The circulation of this function is equal to one
along one edge (in the figure, the vertical edge shared by two adjacent lattice cells) 
and zero along all other edges of the lattice. Within each cell, the interpolation function varies linearly.
A formal expression for four of these functions is

\begin{equation}
   \mathbf{w}_{1-4} \,=\,
   \frac{1}{8 a} \left( 1 \pm \frac{y}{b}\right)
   \left( 1 \pm \frac{z}{c}\right)
   \hat{x}.
\label{eq:walpha}
\end{equation}
Another eight functions of this kind are obtained by the simultaneous cyclic permutation of
$(x,y,z)$ and $(a,b,c)$ in the expression above. For each lattice cell, there are 12 such
interpolating functions altogether (one per edge). Each
function $\mathbf{w}_\alpha$ has unit circulation along edge $\alpha$ ($\alpha = 1,2,\ldots12$)
and zero circulations along all other edges.

The coarse-grained $\bfE$ and $\bfH$ fields can then be represented
by interpolation from the edges into the volume of the cell as follows:
%
%\vskip -0.15in
\begin{equation}\label{eqn:hexahedral-edge-interpolation-EH}
    \bfE ~=~ \sum_{\alpha=1}^{12} \, [\bfe]_\alpha \bfw_\alpha,
    ~~
    \mu_0 \bfH ~=~ \sum_{\alpha=1}^{12} \, [\bfb]_\alpha \bfw_\alpha,
\end{equation}
%\vskip -0.1in
%
\noindent where $[\bfe]_\alpha=\int_\alpha \bfe\cdot \text{d}\mathbf{l}$ is the circulation of the
(microscopic) $\bfe$ field along edge $\alpha$; similar for the $\bfb$ field circulation $[\bfb]_\alpha$. %or $[\bfb]_\alpha$.
In the calculation of the circulations, integration along the edge
is always assumed to be in the positive direction of the respective coordinate axes.

\begin{figure}
  \centering
  \includegraphics[width=8cm]{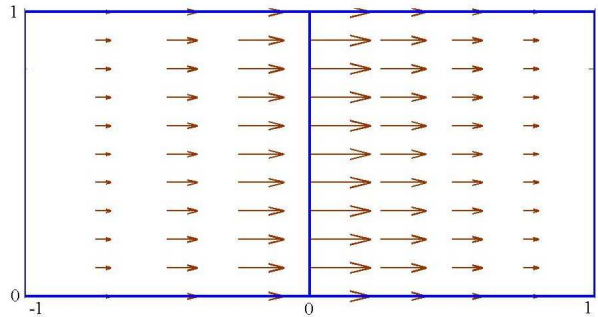}\\ %0.9\linewidth
  \caption{(From \cite{tsukerman}.) A 2D analog of the vectorial interpolation function $\bfv_\beta$
  (in this case, associated with the central vertical edge).
  Normal continuity of this function is evident from the arrow plot;
  its flux is equal to one over the central edge and zero
  over all other edges.}\label{fig:vector-face-function}
\end{figure}

We now move on to the second kind of interpolation that preserves the normal continuity
and produces the $\bfD, \bfB$ fields from $\bfd$ and $\bfb$. A typical interpolating function
(2D rendition again for simplicity) is shown in \figref{fig:vector-face-function}.
The flux of this function through a face shared by two adjacent cells is equal to one;
the flux through all other faces is zero. There are six such functions per cell:
These vector functions can be formally expressed as
\begin{equation}
    \bfv_{1-6} = \left\{ \frac{1}{8bc}\left( 1 \pm \frac{x}{a} \right)\hat{x} \right.\text{,}\left. \frac{1}{8ac}\left( 1 \pm \frac{y}{b} \right)\hat{y} \right. \text{,}
    \left. \frac{1}{8ab}\left( 1 \pm \frac{z}{c} \right)\hat{z} \right\},
\end{equation}
and can be used to define the coarse-grained $\bfD$ and $\bfB$ fields
by interpolation from the six faces into the volume of the unit cell:
\begin{equation}\label{eqn:hexahedral-edge-interpolation-DB}
    \bfD \,=\, \sum_{\beta=1}^{6} [[\mathbf{d}]]_\beta \mathbf{v}_\beta,
    ~~~
    \bfB \,=\, \sum_{\beta=1}^{6} [[\mathbf{b}]]_\beta \mathbf{v}_\beta,
\end{equation}
where $[[\mathbf{d}]]_\beta = \int_\beta{\mathbf{d}\cdot \text{d}\mathbf{S}}$
is the flux of $\bfd$ through face $\beta$ ($\beta = 1,2,\ldots,6$);
similar for the $\bfb$ field. In the calculation of fluxes, it is convenient
to take the normal to any face in the positive direction of the respective coordinate axis
(rather than in the outward direction).

The coarse-grained $\bfE$ and $\bfH$ fields so defined have 12 degrees of freedom in any given lattice cell. 
From the mathematical perspective, these fields lie in the 12-dimensional functional space
spanned by functions $\bfw_{\alpha}$; we shall keep
the notation $W_\mathrm{curl}$ of \cite{tsukerman} for this space: the `W' honors Whitney
\cite{whitney} (see \cite{tsukerman} for details and further references)
and `curl' indicates fields whose curl is a regular function
rather than a general distribution. This implies, in physical terms,
the absence of equivalent surface currents and the tangential continuity of the fields involved.

Similarly, $\bfD$ and $\bfB$ within any lattice cell lie in the six-dimensional
functional space $W_\mathrm{div}$ spanned by functions $\bfv_{\beta}$.
Importantly, it can be shown that the div- and curl-spaces are compatible in the following sense:
\begin{equation}\label{eqn:curl-W-in-div}
    \nabla \times W_\mathrm{curl} \in W_\mathrm{div}
\end{equation}
That is, the curl of any function from  $W_\mathrm{curl}$ (i.e. the curl of
any coarse-grained field $\bfE$ or $\bfH$ defined by \eqnref{eqn:hexahedral-edge-interpolation-EH})
lies in $W_\mathrm{div}$. Because of this compatibility of interpolations,
the coarse-grained fields, as proved in \cite{tsukerman}, satisfy
Maxwell's equations exactly. By construction, they also satisfy the
proper continuity conditions at all interfaces. Thus the coarse-grained
fields are completely physical.

%%%%%%%%%%%%%%%%%%%%%%%%%%%%%%%%%%%%%%%%%%%%%%%%%%%%%%%%%%%%%%%%%
\subsection{Approximation of the Fields and Linear Relationships}\label{sec:Approximation-of-Fields}
%%%%%%%%%%%%%%%%%%%%%%%%%%%%%%%%%%%%%%%%%%%%%%%%%%%%%%%%%%%%%%%%%
%
The vectorial interpolations above are written for the exact fields
that are, obviously, in general unknown. One therefore looks for a good
approximation via a linear combination of basis modes $\psi_\gamma$
\cite{tsukerman}:
$$
    \Psi^{eh} \,=\, \sum\nolimits_\alpha c_\gamma \psi_\gamma^{eh}; ~~~~
    \Psi^{db} \,=\, \sum\nolimits_\alpha c_\gamma \psi_\gamma^{db}.
$$
$\Psi$ and all $\psi_\gamma$ denote, in the most general case, six-component vectors
comprising both microscopic fields; e.g. $\Psi^{eh} \equiv \{ \Psi^e, \Psi^h \}$, etc.;
$c_\gamma$ are complex coefficients.

To each basis wave $\psi_\gamma$, there correspond the curl-interpolants
$\bfE_\gamma(\bfr) = \mathcal{W}_{\mathrm{curl}} ([\psi^{e}_\gamma]_{1-12})$
and $\bfH_\gamma(\bfr) = \mathcal{W}_{\mathrm{curl}} ([\psi^b_\gamma]_{1-12})$.
To define these interpolants (coarse-grained fields), one computes the edge circulations of each
basis mode and applies the interpolation formula \eqnref{eqn:hexahedral-edge-interpolation-EH}.
Similarly, there are the div-interpolants
$\bfD_\gamma(\bfr) = \mathcal{W}_{\mathrm{div}} ([[\psi^d_\gamma]]_{1-6})$
and $\bfB_\gamma(\bfr) = \mathcal{W}_{\mathrm{div}} ([[\psi^b_\gamma]]_{1-6})$,
obtained by computing the face fluxes of the basis modes and then applying
interpolation \eqnref{eqn:hexahedral-edge-interpolation-DB}.

For all basis waves $\alpha$ at any given point $\bfr$ in space, we seek a linear
relation
$$
     \psi_{\gamma}^{DB} (\bfr) ~=~ \bar{\bar{\Omega}}(\bfr) \psi_{\gamma}^{EH} (\bfr).
$$
The $6\times 6$ parameter matrix $\bar{\bar{\Omega}}$ characterizes, in general, anisotropic material behavior with (if the off-diagonal blocks
are nonzero) magnetoelectric coupling. Similar to $\psi^{db}_\gamma$, the six-component vectors $\psi^{DB}_\gamma$
comprise both fields, but coarse-grained; same for $\psi^{EH}_\gamma$.
In matrix form, the above equations are
\begin{equation}\label{eqn:Psi-DB-eq-zeta-Psi-EH}
      \bar{\bar{\Psi}}^{DB} (\bfr) ~=~ \bar{\bar{\Omega}}(\bfr) \bar{\bar{\Psi}}^{EH} (\bfr)  
\end{equation}
where each column of the matrices $\bar{\bar{\Psi}}^{DB}$ and $\bar{\bar{\Psi}}^{EH}$ contains
the coarse-grained fields of the respective basis function.

In this paper, the basis modes $\psi_\gamma$ ($\gamma = 1,2,\ldots,12$)
are chosen as a set of twelve Bloch waves propagating
in six coordinate directions ($\pm x, \pm y, \pm z$), with two different polarizations per direction.
This leads to an overdetermined system of 12 constitutive relations for the $6 \times 6$ parameter
matrix. This system is solved in the least-squares sense. Qualitatively, a good least-squares fit
implies that different modes in the metamaterial can be described accurately enough via
the same set of effective parameters, i.e. the material is ``homogenizable''. A poor fit
corresponds to a ``non-homogenizable'' material; this can also be viewed as strong
``spatial dispersion''. (Rigorous quantitative definitions to be given in \cite{tsukerman-inprep}).

Mathematically, the material parameter matrix is appropriately expressed
via the pseudoinverse \cite{allaire}:
\begin{equation}
    \bar{\bar{\Omega}}(\mathbf{r}) \,=\,
    \bar{\bar{\Psi}}^{DB}(\mathbf{r})\left(\bar{\bar{\Psi}}^{EH}(\mathbf{r})\right)^+.
\label{eq:parammatrix}
\end{equation}
This expression is equivalent to solving the least-squares problem.

The parameter matrix $\bar{\bar{\Omega}}(\mathbf{r})$, being a function
of position $\mathbf{r}$, is an interim quantity. The effective parameters are found by averaging
$\bar{\bar{\Omega}}(\mathbf{r})$ over the whole unit cell \cite{tsukerman}.
This accomplishes our goal of finding an effective $6 \times 6$ material matrix
that describes, in the frequency domain,
linear constitutive relations between the different electromagnetic fields:
\begin{equation}
   \begin{pmatrix}
      \mathbf{D}(\mathbf{r},\omega) \\ \mathbf{B}(\mathbf{r},\omega)
   \end{pmatrix}
   =
   \begin{pmatrix}
      \bar{\bar{\varepsilon}}(\omega) &  \bar{\bar{\xi}}(\omega) \\
      \bar{\bar{\zeta}}(\omega) & \bar{\bar{\mu}}(\omega)
   \end{pmatrix}
\begin{pmatrix} \mathbf{E}(\mathbf{r},\omega) \\ \mathbf{H}(\mathbf{r},\omega) \end{pmatrix}.
\label{eq:relation}
\end{equation}
The frequency-dependent parameter matrix $\bar{\bar{\Omega}}$ comprises four $3\times 3$ matrices with $\bar{\bar{\varepsilon}}$
and $\bar{\bar{\mu}}$ describing the electric and magnetic response, respectively, whereas $\bar{\bar{\xi}}$
and $\bar{\bar{\zeta}}$ characterize the magnetoelectric coupling.

When accurate analytical approximations of the microscopic fields are
available, they may be used to generate the basis set; however,
this is feasible only in the simplest cases. Generally, numerical techniques have to be employed.
In this paper, the finite element method (FEM) is used to find the basis modes in the cell.
As noted above, these modes are chosen as twelve Bloch waves of the form
$\mathbf{e} = \mathbf{e}_{per}\exp(ik_B \hat{\mathbf{n}}_{dir} \cdot \mathbf{r})$
(with the phasor convention $\exp[-i\omega t]$). Here $\mathbf{e}_{per}$ is a periodic amplitude function,
$k_B$ is the complex Bloch wavenumber and $\hat{\mathbf{n}}_{dir}$ 
is the unit vector of propagation direction.
If the Bloch wave solution is substituted into the vector wave equation, one obtains an eigenvalue equation for $\bfe_{per}$
\begin{equation}
    \left(\nabla+ik_B \hat{\mathbf{n}}_{dir}\right)
    \times \left(\nabla+ik_B \hat{\mathbf{n}}_{dir}\right)
    \times \mathbf{e}_{per} - k_0^2\epsilon \mathbf{e}_{per}
    \,=\, 0,
\label{eq:wave}
\end{equation}
where $k_B$ is an eigenvalue and $\mathbf{e}_{per}$ is an eigenmode.
In this paper, this eigenvalue solver has been implemented in the commercial FEM software
Comsol Multiphysics \cite{urzhumov}. Since the eigenproblem is written for the periodic factor $\mathbf{e}_{per}$
rather than the total field, periodic boundary conditions are applied on all three pairs
of faces of the unit cell \cite{davanco}. By considering six directions of propagation $\hat{\mathbf{n}}_{dir}$
and two polarizations per direction, one constructs a 12-mode basis for the homogenization procedure. The construction of the 12-mode basis requires for each frequency three independent eigenvalue calculations corresponding to $\hat{\mathbf{n}}_{dir}$ being chosen along the three coordinate axes. The reason why only three calculations are sufficient (and not six) is because the eigenvalue solver computes the $\pm k_B$ solutions, i.e., each choice of $\hat{\mathbf{n}}_{dir}$ covers the two propagation directions $\pm \hat{\mathbf{n}}_{dir}$. As an example, if $\hat{\mathbf{n}}_{dir}=(1,0,0)$ the basis modes for $\pm x$-propagation with $y$- and $z$-polarization are obtained (four basis modes are determined per $\hat{\mathbf{n}}_{dir}$). 
Note that, in addition to propagating modes, the calculation may reveal some evanescent ones.
These modes do not affect the optical properties of voluminous bodies but may be considered
in future studies where surface waves are of interest.
It goes without saying that, when the edge circulations and face fluxes 
are calculated for the construction of the coarse grained fields, 
the microscopic field is not, e.g., $\mathbf{e}_{per}$ but rather 
$\mathbf{e}_{per}\exp(ik_B \hat{\mathbf{n}}_{dir} \cdot \mathbf{r})$.

\section{Test cases} \label{sec:testcases}
This section is divided into three subsections. The first one discusses the seemingly
trivial, and yet instructive, case of a homogeneous cell. In the second subsection,
all 36 effective parameters are computed for a metamaterial with spherical gold nanoparticles;
to test the methodology, no \emph{a priori} symmetry assumptions are built into the procedure.
In the third subsection, a metamaterial with gold SRRs is considered and all 36 parameters
are again found without any preconditions. In all our simulations, the Johnson \& Christy
data \cite{johnson} for the dielectric function of gold are used.

\subsection{Case 1: The homogeneous cell revisited}
Effective parameters of a homogenous cell, as a ``sanity check'' of the proposed method, 
were already discussed in \cite{tsukerman}. It was shown that the method
indeed yields the exact results, i.e., $\bar{\bar{\varepsilon}}$ and $\bar{\bar{\mu}}$
are diagonal matrices with the values equal to the intrinsic permittivity and permeability
of the homogeneous material ($\epsilon$ and $\mu$, respectively), and $\bar{\bar{\xi}}$ and $\bar{\bar{\zeta}}$ are zero matrices 
(no magnetoelectric coupling). Although this result appears trivial, we note that
in some existing theories spurious Bloch-like factors do appear in the effective parameters
and then need to be discarded by fiat.

To elaborate, the homogenization procedure outlined above yields for a uniform cell,
after a direct analytical calculation, the effective permittivity
\begin{equation}
    \varepsilon_{ii} 
    \,=\,
    \epsilon \frac{\arctan(\tan(k a))}{k a},
    ~~~ k = \omega \sqrt{\mu \epsilon}
\label{eq:epsilon}
\end{equation}
which results in $\varepsilon_{ii} \,=\, \epsilon$
if $k a<\pi/2$; the subscript $ii$ indicates that only diagonal elements are different from zero.
Thus the method is applicable when the wavelength is greater than double the unit cell size
($4a<\lambda$). This is perfectly consistent with the fact that homogenization
makes sense only when the unit cell is substantially subwavelength.

\subsection{Case 2: Spherical gold particle}
Effective parameters of a material with subwavelength-sized spherical particles
are described well by the Maxwell-Garnett formula \cite{maxwell,doyle} and the Lewin theory
for a cubic lattice of spheres \cite{lewin}. This is therefore an excellent test case,
where the new procedure can be tested against the known analytical results
and the dispersion around the plasmon resonance wavelength is very strong,
a property that any useful homogenization method must capture.

In the following, a spherical gold particle with radius $r_0=20$~nm
in a cubic unit cell of half-size $a=b=c=40$~nm is considered.
The host material is assumed to be non-polarizable (or free space).
As the volume fraction of the spheres is only $v_f \simeq 0.07$ and the wavelength interval of consideration is $300-900$~nm, the calculated results
can appropriately be compared with the Lewin expression for the diagonal entries of
the effective permittivity matrix \cite{lewin}:
\begin{equation}
\label{eq:epslewin}
   \varepsilon_{ii}=\epsilon_h \left( 1+\frac{3 v_f}{\frac{F(\theta)+2\epsilon_h /\epsilon_{s}}{F(\theta)-\epsilon_h/\epsilon_{s}}-v_f} \right),
\end{equation}
where
$F(\theta) = 2(\sin\theta-\theta\cos\theta) /
[(\theta^2-1)\sin\theta+\theta\cos\theta]$,
$\theta = k_0 r_0 \sqrt{\epsilon_{s}\mu_{s}/(\epsilon_0\mu_0)}$,
$\epsilon_{s}$ and $\mu_{s}$ the permittivity and permeability of the spheroidal inclusions,
respectively, and $\epsilon_h$ the permittivity of the host material.
For gold spheres in a vacuum, $\epsilon_{s} = \epsilon_{AU}(\omega)$,
$\mu_{s}=\mu_0$ and $\epsilon_h = \epsilon_0$. Subscript \textit{ii} in Eq. (\ref{eq:epslewin})
refers to either $xx$, $yy$ or $zz$. The permittivity tensor is obviously diagonal;
however, this is not explicitly built into the homogenization procedure as an \emph{a priori}
assumption but rather is expected to emerge naturally as a result.
Also, the Lewin theory does not include any magnetoelectric coupling ($\bar{\bar{\xi}}$
and $\bar{\bar{\zeta}}$ are zero matrices), whereas the magnetic response is described
as in Eq. (\ref{eq:epslewin}) with the substitution of $\epsilon$ with $\mu$.

\begin{figure}
\centering
		\includegraphics[width=8cm]{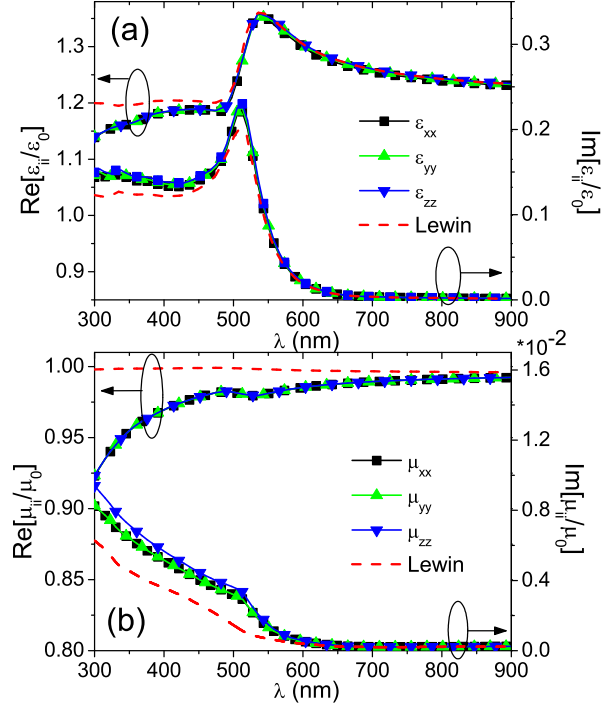}
	\caption{The diagonal elements of a) the effective permittivity matrix and b) permeability matrix for a metamaterial consisting of cubic-ordered gold spheres with diameter of 40nm embedded in a vacuum. The unit cell size is (80nm)$^3$. The ovals and arrows identify which curves belong to which axes.}
	\label{fig:sphereparam}
\end{figure}

The procedure of \cite{tsukerman} and Sec. \ref{sec:method} has yielded all 36 parameters
that agree quantitatively well with the Lewin theory (Fig. \ref{fig:sphereparam}).
In this figure, only the diagonal elements of the permittivity and permeability matrices
are displayed. All other computed parameters are negligibly small, although not identically zero
due to numerical errors. Specifically, the off-diagonal elements of $\bar{\bar{\varepsilon}}$
and $\bar{\bar{\mu}}$ are $10^4-10^6$ times smaller than the diagonal elements,
whereas the dimensionless values of $c_0\bar{\bar{\xi}}$ and $c_0 \bar{\bar{\zeta}}$
($c_0$ is the speed of light in  a vacuum) are smaller by a factor of $\sim 10^5$ compared to the values
of $\bar{\bar{\varepsilon}}/\varepsilon_0$ and $\bar{\bar{\mu}}/\mu_0$,
thereby confirming that the effect of the magnetoelectric coupling is negligible.

The computed diagonal elements of the permittivity matrix are all equal and also
consistent with the Lewin theory [Fig. \ref{fig:sphereparam}(a)], exhibiting only a small discrepancy
at shorter wavelengths. Even the plasmon resonance at $\simeq 520$~nm, resulting in a strong dispersion
and increased absorption, is well captured in the effective parameters.
The higher discrepancy at shorter wavelengths is a natural consequence of approaching
the homogenization limit (double the cell size) at approximately $\lambda \sim 160$~nm. The permeability matrix [Fig. \ref{fig:sphereparam}(b)]
is also consistent with the expected behavior; its diagonal elements are predominantly real
and equal to $\approx 1$.

The agreement with the Lewin theory is, however, not perfect. At shorter wavelengths,
the real part of the permeability is smaller than Lewin's value of $\sim 1$,
while the small imaginary part is slightly overestimated as compared to Lewin's theory.
These discrepancies are a result of leaving the subwavelength regime.
One should note that although the relative deviation is larger for $\mu_{ii}$ compared to
$\varepsilon_{ii}$, the absolute difference is approximately $\sim 0.1$ for both $\varepsilon_{ii}$
and $\mu_{ii}$. It is also worth noting that the retrieved permeability is affected to some extent
by the electric plasmon resonance and exhibits a weak perturbation near the resonance wavelength.
We believe this is a consequence of numerical errors.
For instance, the $\mathbf{b}$ field is obtained
by numerical differentiation from the calculated electric field in the eigenvalue
problem [see Eq. (\ref{eq:wave})] which affects the accuracy of the respective
edge circulations and face fluxes. Importantly, though, both $\bar{\bar{\varepsilon}}$
and $\bar{\bar{\mu}}$ satisfy the passivity requirement
$\text{Im}[\varepsilon_{ii}]>0$ and $\text{Im}[\mu_{ii}]>0$.

\subsection{Case 3: Split-ring resonators} \label{sec:srr}
\begin{figure}
	\centering
		\includegraphics[width=8cm]{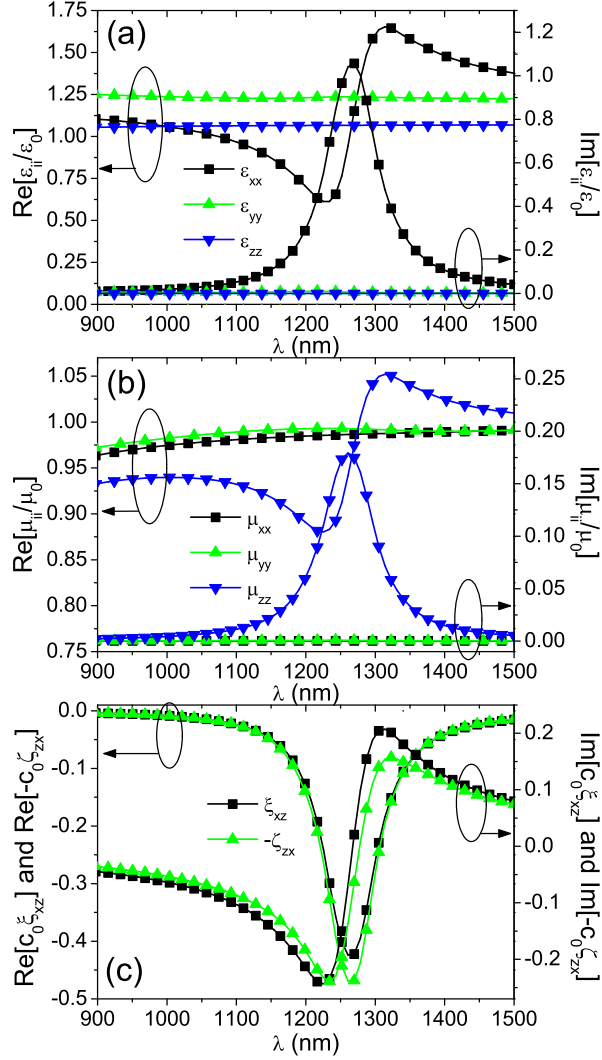}
	\caption{The dominant elements of a) the effective permittivity matrix,
      b) effective permeability matrix, and c) the magnetoelectric coupling matrices
      for a metamaterial consisting of cubic-ordered gold SRRs of Fig.~\ref{fig:setup}.
      The unit cell size is (200nm)$^3$. The ovals and arrows identify which curves belong to which axes.}
\label{fig:srrparam}
\end{figure}

The split-ring resonator (SRR) is the ubiquitous subwavelength meta-atom
in metamaterial design \cite{linden}, giving rise to artificial magnetism
or negative refractive index \cite{shelby}. Even though SRR is generally utilized
as a magnetic meta-atom, it also has, as noted in e.g. \cite{chen,li},
a magnetoelectric response characteristic of optical activity. In the following,
a gold C-shaped SRR in an otherwise empty cubic unit cell with $a=b=c=100$nm is analyzed.
Having a width and thickness of 35~nm, the SRR lies in the $xy$-plane parameterized by a radius of 35~nm and a gap of 9~nm [Fig. \ref{fig:setup}(b)].
With this configuration, the magnetoelectric response arises due to an $x$-polarized
incident wave inducing a magnetic dipole moment along the $z$-axis, and, conversely,
a magnetic field along the $z$-axis generating an electric dipole moment along the $x$-axis.
Thus the effective parameter matrices take on the following form:
\begin{align}
& \bar{\bar{\varepsilon}}=\begin{pmatrix} \varepsilon_{xx} & 0 & 0 \\ 0 & \varepsilon_{yy} & 0 \\ 0 & 0 & \varepsilon_{zz} \end{pmatrix}
& \bar{\bar{\mu}}=\begin{pmatrix} 1 & 0 & 0 \\ 0 & 1 & 0 \\ 0 & 0 & \mu_{zz} \end{pmatrix} \nonumber \\
& \bar{\bar{\xi}}=\begin{pmatrix} 0 & 0 & \xi_{xz} \\ 0 & 0 & 0 \\ 0 & 0 & 0 \end{pmatrix}
& \bar{\bar{\zeta}}=\begin{pmatrix} 0 & 0 & 0 \\ 0 & 0 & 0 \\ \zeta_{zx} & 0 & 0 \end{pmatrix},
\label{eq:srrmatrices}
\end{align}
where $\xi_{xz}=-\zeta_{zx}$ due to reciprocity \cite{marques}.
As in the previous test of the new methodology, no advance information about the
symmetry and structure of the parameter matrices was explicitly utilized.
However, as in the previous case, only the dominant parameters
are shown in Fig. \ref{fig:srrparam}.%Eq. (\ref{eq:srrmatrices}).

The off-diagonal elements in $\bar{\bar{\varepsilon}}$ and $\bar{\bar{\mu}}$ are $10^4-10^6$ times
smaller than the diagonal elements, while the elements in $\bar{\bar{\xi}}$ and $\bar{\bar{\zeta}}$
are smaller than $\xi_{xz}$ and $\zeta_{zx}$ by a factor of $10^3-10^5$.

It is evident from Fig. \ref{fig:srrparam}(a) that $\varepsilon_{xx}$ exhibits a plasmonic resonance response
at $\simeq 1260$~nm. This is due to the induced electric dipole moment pointing along the gap,
in the $x$-direction. In contrast, $\varepsilon_{yy}$ and $\varepsilon_{zz}$ are
almost frequency-indpendent, representing just a perturbation of the unit parameters
of the empty cell by the inclusion of gold SRRs. Concurrently with the electric response,
the SRR-metamaterial also displays a magnetic response in $\mu_{zz}$ [Fig. \ref{fig:srrparam}(b)],
dramatically increasing the imaginary part while the real part shows diamagnetic
and paramagnetic behavior below and above the resonance wavelength, respectively.
This is all in accordance with analytical approximations of SRR by an equivalent circuit
diagram \cite{cai}. Again, it is noticeable how the non-resonant diagonal elements
decrease slightly ($\mu_{ii}/\mu_0<1$) for shorter wavelengths (limit of homogenization is at $\sim 400$~nm).

For easier comparison of the the dimensionless magnetoelectric coupling parameters,
Fig. \ref{fig:srrparam}(c) shows $c_0\xi_{xz}$ and $-c_0\zeta_{zx}$.
It is evident that $\xi_{xz}$ and $-\zeta_{zx}$ are very similar but not identical;
there are noticeable deviations from reciprocity near the resonance.
We believe that the small difference in the parameters, $c_0|\xi_{xz}+\zeta_{zx}|$,
indicates the approximation error inherent in the procedure.
At the resonance, where the difference is largest ($c_0|\xi_{xz}+\zeta_{zx}|\simeq 0.1$),
we expect the approximation error to be higher.

\section{Comparison with the S-parameter method} \label{sec:compare}
Even though the computed parameters exhibit the expected behavior for 
both the sphere- and SRR-metamaterial,
there is no analytical result in the plasmonic regime for the SRR-metamaterial that can be used for direct comparison (only approximations derived from electric circuit models are available \cite{marques,cai,pendry}).
In order to quantitatively verify the calculated effective parameters from Sec. \ref{sec:srr}, this section studies and compares parameters of the SRR-metamaterial obtained with
the well-established S-parameter method originally proposed
for isotropic metamaterials \cite{smith} and then extended to anisotropic and bianisotropic
metamaterials of the SRR-type \cite{chen}. In this paper, a simplified retrieval procedure
of Ref. \cite{li} is used. In these procedures, regardless of their specific variants,
the effective parameters are data-fitted to the complex reflection and transmission coefficients
of a finite-sized slab of the metamaterial. In our calculations, the SRR-slab has the thickness
of three unit cells (600~nm). In the following we denote the S-parameter method `S' or the `S-method'.

The method studied in this paper will be denoted `D' or `D-method' as an abbreviation of `Direct', thereby emphasizing the fact that effective parameters are derived from the linear relations between coarse-grained fields. To bring the conventions
in line with the S-parameter approach, in which only a subset of the effective parameters can be retrieved, the magnetoelectric coupling is described
by a single parameter $\chi_0$ in which $\xi_{xz}=-i\chi_0$ and $\zeta_{zx} = i\chi_0$.
Hence, for the D-method the parameter is defined as $\chi_0 = i(\xi_{xz}-\zeta_{zx})/2$.

\begin{figure*}
	\centering
		\includegraphics[width=1\textwidth]{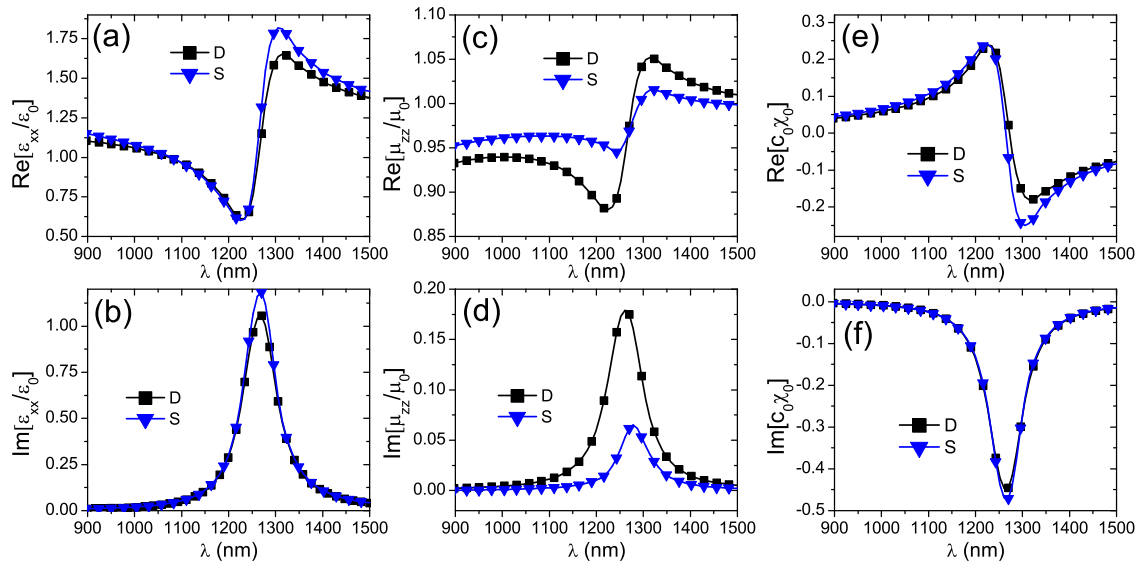}
	\caption{Comparison of the parameters $\varepsilon_{xx}$, $\mu_{zz}$ and $\chi_0$ for the SRR-metamaterial obtained by the direct method (`D') presented in this paper and the S-parameter method (`S') in Ref. \cite{li}.}
	\label{fig:srrcompare}
\end{figure*}

The SRR-metamaterial is described by five effective parameters in the S-method, but for comparison it is only interesting to study $\varepsilon_{xx}$, $\mu_{zz}$ and $\chi_0$,
as they are the only parameters affected by the plasmon resonance.
The parameters are displayed in Fig \ref{fig:srrcompare}.
The electric response
[Fig. \ref{fig:srrcompare}(a)-(b)] is very similar for both methods,
with only a small variation near the resonance.
In contrast, the difference is substantial in the magnetic response [Fig. \ref{fig:srrcompare}(c)-(d)]
with $\text{Im}[\mu_{zz}]$ exhibiting a difference of a factor of $\sim 2.5$.
Similarly, $\text{Re}[\mu_{zz}]$ has noticeable discrepancies, with both methods showing a slight decrease in $\mu_{zz}$ at shorter wavelengths. 
The situation is somewhat different for the magnetoelectric coupling [Fig. \ref{fig:srrcompare}(e)-(f)],
where the two methods show curves close to one another that from physically intuition behave as expected with a response of $\text{Re}[c_0\chi_0]$ situated around $\text{Re}[c_0\chi_0]=0$.

\begin{figure}
	\centering
		\includegraphics[width=8cm]{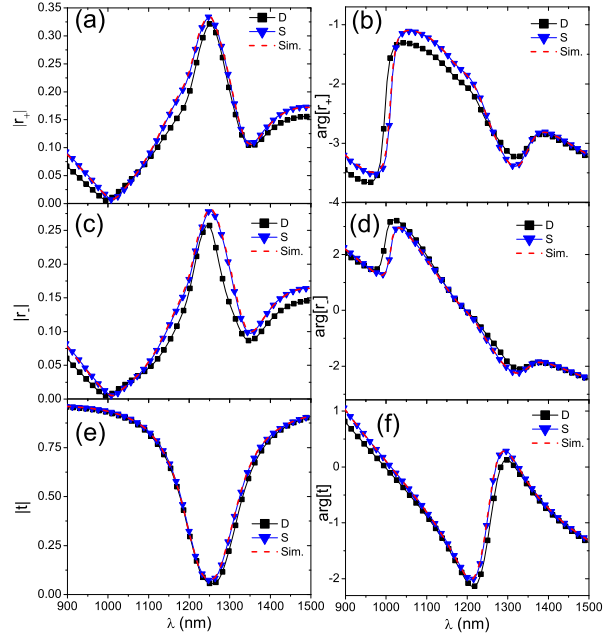}
	\caption{Comparison of the complex reflection and transmission coefficients from a 1$\mu$m thick SRR-metamaterial slab for the two different homogenization methods. The incident wave is x-polarized and propagating along the y-direction. $r_+$ and $r_-$ correspond to the reflection coefficients for propagation along +y and -y, respectively. The legend `Sim' denotes full-wave simulation of a 5-layer SRR-metamaterial.}
	\label{fig:srrtr}
\end{figure}

It is evident from the comparison above that both methods qualitatively agree well,
but the specific values differ in some instances. In relation to the transmission
and reflection coefficients, these parameter differences might conceivably add up or perhaps
counterbalance each other, thereby leading to less or more similar spectra.
To investigate this further, the reflection and transmission coefficients ($r_\pm$ and $t$)
for a $1\mu$m thick SRR-metamaterial slab are calculated for a normally incident
$x$-polarized plane wave propagating in the $y$-direction (Fig. \ref{fig:srrtr})
using the expressions \cite{li}
\begin{align}
\label{eq:reflek}
   & r_\pm=  \frac{2i\sin(n k_0 l)[n^2+(\chi_0^r\pm i\mu_{zz}^r)^2]}{[(\mu_{zz}^r+n)^2+(\chi_0^r)^2]e^{-in k_0 l} -[(\mu_{zz}^r-n)^2+(\chi_0^r)^2]e^{in k_0 l}}\\
   & t=  \frac{4\mu_{zz}^r n}{[(\mu_{zz}^r+n)^2+(\chi_0^r)^2]e^{-in k_0 l} -[(\mu_{zz}^r-n)^2+(\chi_0^r)^2]e^{in k_0 l}},
\label{eq:trans}
\end{align}
where $l$ is the thickness of the slab and $n = \sqrt{\epsilon_{xx}^r\mu_{zz}^r-(\chi_0^r)^2}$
the refractive index of the metamaterial. The superscript $r$ is used to emphasize
that the effective parameters are the relative dimensionless values ($\mu_{zz}^r = \mu_{zz}/\mu_0$ etc.).
Similarly, the subscript $\pm$ on the reflection coefficient indicates that the reflection
is sensitive to the $\pm y$-direction of incidence due to strong bianisotropy
for the chosen direction and polarization. One should also note that the effective parameters
of the S-method are in fact retrieved from the transmission and reflection coefficients
in Eq. (\ref{eq:reflek})-(\ref{eq:trans}) for a 600-nm thick SRR-metamaterial slab.
For this reason, the spectra with parameters from the S-method are expected to be in (almost) perfect agreement with the full-wave simulations of the transmission and reflection spectra. 
This is indeed also the case, as evident from Fig. \ref{fig:srrtr}, where the curves from the S-method and full-wave simulation are coinciding. Consequently, it can be deduced that a 3-layer SRR-metamaterial is sufficient to retrieve the "bulk" effective parameters, i.e., the effective parameters of the S-method (Fig. \ref{fig:srrcompare}) are converged. 

The comparison of the D-method spectra with the full-wave simulations are equivalent to comparing the two homogenization approaches. Comparing the reflection spectra obtained by the two methods [Fig. \ref{fig:srrtr}(a)-(d)], it is remarkable how well the curves are consistent with one another, even though the retrieved permeability $\mu_{zz}$ is significantly different
[Fig. \ref{fig:srrcompare}(c)-(d)]. The difference in effective parameters seems to (partially) counterbalance each other. 
That said, it is still clear that the reflection weakly depends on the choice of effective parameters; this dependence is, however, much less rigid for transmission. The transmission coefficient
[Fig. \ref{fig:srrtr}(e)-(f)] is only marginally affected by the choice of the method,
with the two methods showing an almost perfect agreement in both the amplitude and phase.
It should be noted that such an insensitivity to the transmission coefficient
is in fact a problem for the S-method, as small numerical errors
in the S-parameters may substantially influence the retrieved effective parameters.

A similar comparison of effective parameters for the SRR-metamaterial
have also been conducted for the SRR-metamaterial with a smaller unit cell half-size $a=b=c=80$nm (not shown).
Naturally, the dispersion of the effective parameters is in this case higher,
but the differences in the parameters between the two methods follow the same trend
as for the larger unit cell (Fig. \ref{fig:srrcompare}). Similarly, the effective parameters of the D-method accurately describe the optical properties of the SRR-metamaterial showing only small deviations in the transmission and reflection spectra compared to the full wave simulations.

\section{Conclusion} \label{sec:conclusion}
The paper describes a general three dimensional implementation of the homogenization theory
developed in \cite{tsukerman}. In contrast with the common ``retrieval'' procedures
where the effective medium parameters are inferred from transmission and reflection
data, the proposed approach follows directly from the definition of material parameters
as linear relations between the coarse-grained fields. To put it differently, our theory is aimed 
at understanding the``inner workings'' of the metamaterial, whereas S-parameters characterize 
it as a black box from the point of view of an external observer. These two categories of methods 
-- ``\emph{ab initio}'' analysis vs. indirect inference -- complement one another. However, since these categories are philosophically different, their comparison should go 
beyond pure utilitarian advantages that one of them may or may not have in a particular case
at a particular stage of development.

The key physical insight in our theory is that the $\bfE$ and $\bfH$ fields must satisfy
the interface continuity conditions different from that of the $\bfD$ and $\bfB$ fields,
and hence two different interpolation procedures are called for.
These interpolations are effected by two types of linear vector functions 
(\sectref{sec:Approximation-of-Fields} and \cite{tsukerman}).
The fact that the $\bfH$ and $\bfB$ fields are subject to different kinds of interpolation
explains the paradox of ``artificial magnetism:'' why the macroscopic $\bfB$ and
$\bfH$ fields can be different even though the underlying microscopic field is the same.

The method and its implementation do not require any \emph{a priori} assumptions about the $6 \times 6$
constitutive matrix and allow one to find all 36 constitutive parameters. 
From the computational perspective, the procedure involves, at any given frequency,
the calculation of 12 Bloch modes whose fields on the faces and edges of the unit cell
are used to construct the coarse-grained fields by the two types of vectorial interpolation.
The result is unambiguously defined coarse-grained fields obeying Maxwell's equations
and the continuity conditions.

Calculations for a homogeneous cell show that the method produces the exact result,
with no spurious factors, as long as the cell size does not exceed one half of the wavelength. 
This condition must always be satisfied in practice, for the homogenization to be meaningful.

The homogenization of a metamaterial with spherical gold particles
has been carried out and compared with the Lewin theory showing, in general, an almost perfect match.
It is only at shorter wavelengths (the unit cell larger than one quarter of the free space wavelength)
that deviations occur and the permeability is estimated to be slightly below unity.

The effective parameters computed for a periodic SRR-metamaterial exhibit
the expected physical bianisotropic behavior with magnetic and magnetoelectric responses.
Furthermore, the parameters are compared with S-parameter retrieval 
exhibiting a qualitatively similar behavior but some quantitative differences.
It was then demonstrated how the differences in parameters influence
the reflection and transmission coefficients of an SRR-metamaterial slab.
It turns out that the new homogenization method describes reflection and transmission
very accurately.

We believe that our successful implementation of the method
will be useful in the future studies and design of complex nanostructures,
especially if the specific properties of the effective material matrix are not known.
Just like the S-parameter procedure \cite{fokin}, the new method and its implementation
may also be transferred to the field of acoustics and applied to acoustic metamaterials.

\begin{acknowledgments}
The authors gratefully acknowledge the help of Yaroslav Urzhumov
(Duke University) in implementing the electromagnetic Bloch mode solver in Comsol Multiphysics.
We also appreciate the useful discussions with Morten Willatzen (University of Southern Denmark).
This work was supported by the Danish Council for Independent Research (FTP-project ANAP, contract No. 09-072949).

I.T. thanks Vadim Markel and Boris Shoykhet for very helpful discussions.
\end{acknowledgments}

%\bibliography{pre_HomogenizationPlasmonicMetamaterials}

\end{document}